\documentclass[aps,prd,twocolumn,amsmath,amssymb,floatfix,superscriptaddress,showpacs]{revtex4-1}
\usepackage{graphicx}
\usepackage{amssymb}
\usepackage{amsmath}
\usepackage{color}
\usepackage{ wasysym }
\usepackage{hyperref}
 \usepackage{tabu}

\def\barray{\begin{array}}
\def\earray{\end{array}}
\def\be{\begin{equation}}
\def\ee{\end{equation}}
\def\ben{\begin{equation} \nonumber}
\def\een{\end{equation}}
\def\ban{\begin{eqnarray*}}
\def\ean{\end{eqnarray*}}
\def\ba{\begin{eqnarray}}
\def\ea{\end{eqnarray}}

\def\({\left(}
\def\){\right)}

\graphicspath{{./fig/}}

\begin{document}

%\title{Gauge-Quintessence}
\title{Growth of matter perturbations in clustered holographic dark energy cosmologies}
\author{Ahmad Mehrabi}
\email{Mehrabi@basu.ac.ir}
\affiliation{Department of Physics, Bu-Ali Sina University, Hamedan
65178, 016016, Iran}

\author{Spyros. Basilakos}
\email{svasil@academyofathens.gr}
\affiliation{Academy of Athens, Research Center for Astronomy \& Applied \\
  Mathematics, Soranou Efessiou 4, 11-527, Athens, Greece }
\email{svasil@academyofathens.gr}

\author{Mohammad Malekjani}
\email{malekjani@basu.ac.ir}
\affiliation{Department of Physics, Bu-Ali Sina University, Hamedan
65178, 016016, Iran}

\author{Zahra Davari}
\affiliation{Department of Physics, Bu-Ali Sina University, Hamedan
65178, 016016, Iran}

\begin{abstract}
We investigate the growth of matter fluctuations in holographic dark energy
cosmologies. First we use an overall statistical analysis
involving the latest observational data in order
to place constraints on the cosmological parameters. Then we test the range of
validity of the holographic dark energy models at the
perturbation level and its variants from the concordance $\Lambda$
cosmology. Specifically, we provide a new analytical approach
in order to derive, for the
first time, the growth index of matter perturbations.
Considering a homogeneous holographic dark energy
we find that the growth index is
$\gamma \approx \frac{4}{7}$ which is somewhat
larger ($\sim 4.8\%$) than that of the usual
$\Lambda$ cosmology, $\gamma^{(\Lambda)}\approx \frac{6}{11}$.
Finally, if we allow clustering in the holographic dark energy models then
the asymptotic value of the growth index is given in terms
of the effective sound speed $c_{\rm eff}^2$, namely $\gamma \approx \frac{3(1-c_{\rm eff}^2)}{7}$.
\end{abstract}
\maketitle

\section{Introduction}
The current accelerated expansion of the Universe
\cite{Riess:1998cb,Perlmutter:1998np,Kowalski:2008ez,Hinshaw:2008kr} can be
well explained either by introducing
the so called dark energy (hereafter DE), namely
an exotic cosmic fluid with a negative pressure,
or by modifying the standard theory of
gravity
on extragalactic scales. Based on the Planck
results \cite{Ade:2013zuv} it has been found that
that DE amounts to $\sim 69\%$ while
the matter component (cold dark matter+baryons)
corresponds to $\sim 31\%$ of the current total energy budget, respectively.
%Although many theoretical and observational studies have been done
%but the nature of DE is still obscure.

In this framework, a large family of cosmological models have been
proposed in order to explain the origin of the cosmic acceleration
(for a comprehensive review see \cite{Copeland:2006wr}).
The simplest DE candidate is the well known
cosmological constant for which the equation of state (hereafter EoS)
parameter is strictly equal to -1 and thus the DE is connected with the
energy of vacuum. Although the concordance $\Lambda$ cosmology is consistent
with the available observational data it has two weak points:
the fine-tuning and the cosmic coincidence %problems
\cite{Sahni:1999gb,Weinberg:1988cp,Carroll:2000fy,Peebles:2002gy,Padmanabhan:2002ji,Copeland:2006wr}. Alternatively, in the last two decades, a wealth of
dynamical DE models with a time-varying EoS have been studied in the literature
\cite{Copeland:2006wr,Li:2011sd,Bamba:2012cp} in order to circumvent
the above cosmological puzzles. However, the majority of DE models are based
on purely phenomenological arguments.

In this work we focus on the holographic dark energy
(HDE) scenario which has a strong theoretical motivation. Indeed
the HDE model originates from the fundamental
holographic principle in quantum gravity theory
\cite{Susskind:1994vu}. In particular, according to the holographic
principle, the number of degrees of freedom for a finite-size system
is finite and bounded by the area of its boundary
\cite{Cohen:1998zx}. Therefore, applying the holographic principle in
cosmology, namely considering the future event horizon for
IR cut-off one can show that the energy density of DE is given by
(see Refs. \cite{Ng:2000fq,Arzano:2006wp,Horava:2000tb,Cataldo:2001bn,Thomas:2002pq,Hsu:2004ri,Li:2004rb})
\begin{equation}\label{eqn:h1}
\rho_{\rm d}=3s^2M_{Pl}^2R_{\rm h}^{-2}\;,
\end{equation}
where $s$ is a constant, $M_{Pl}^2=1/8 \pi G$ is the
reduced Plank mass and the coefficient $3$ is just for convenience.
Notice, that the event horizon is written as
\begin{equation}\label{eqn:h2}
 R_{\rm h}= a\int^{\infty}_t{\frac{dt}{a(t)}}=a\int_a^{\infty}{\frac{da}{a^2 H(a)}}\;,
\end{equation}
where $a(t)$ is the scale factor of the universe, $H(a)=\frac{\dot a}{a}$ is the Hubble
parameter and $t$ is the cosmic time. Interestingly,
it has been found that the HDE cosmological pattern provides the current
cosmic acceleration and it is in agreement with
the observational data
\cite{Pavon:2005yx,Zimdahl:2007zz,Sheykhi:2011cn,Huang:2004wt,Kao:2005xp,Zhang:2005hs,Wang:2005ph,Chang:2005ph,
Zhang:2007sh,Micheletti:2009jy,Xu:2012aw,Zhang:2013mca,Li:2013dha,Zhang:2014ija,Zhang:2015rha}. Moreover, in this scenario the coincidence as well as the
fine-tuning problems are successfully alleviated
\cite{Li:2004rb}.

However, in addition to the background evolution, the formation of large
scale structures provides valuable information about the nature of DE
\cite{Tegmark:2003ud}. Indeed, matter perturbations can grow via
the gravitational instability during the different epochs of the
cosmic history.
%starting from inflationary era up to current DE
%dominated phase.
In fact, DE component not only accelerates the expansion
of the Universe but also it affects the growth rate of matter
perturbations \cite{Peebles1993}.
It becomes clear that the latter opens a new avenue
towards understanding the mechanism of structure formation in the DE
regime.

From the observational viewpoint, it
is worthwhile to set up a more general formalism in which the
background data (SnIa, BAOs, CMB shift parameter etc)
are jointed to the growth data (Ref.\cite{Mehrabi:2015hva} and references therein) in 
order to place constraints on the DE models.
Concerning the HDE models such a joint analysis has given
$s=0.750_{-0.0999}^{+ 0.0976}$ and $\sigma_8=0.763_{- 0.0465}^{+0.0477}$
\cite{Xu:2013mic}, where $\sigma_8$ is
the variance of matter perturbations within the sphere of
$R=8{\rm h^{-1}Mpc}$ at present time.
It is worth mentioning that the
author of \cite{Xu:2013mic} has treated the holographic DE
as homogeneous.

Generally speaking, in the case of dynamical DE models, one can consider
fluctuations in both time and space in a similar fashion to matter
\cite{Abramo:2007iu,Abramo:2008ip,Batista:2013oca,Mehrabi:2014ema}.
Potentially, since the energy density of HDE 
is defined according to event horizon IR cut-off, one may consider that 
the origin of the HDE perturbations is due to the fluctuations of the future 
event horizon \cite{Li:2008zq}. In this case, it has been shown that the HDE adiabatic sound speed  
is positive and therefore the corresponding perturbations are stable (see \cite{Li:2008zq}).
%Since it is difficult to include the HDE component into a Lagrangian 
%formulation, initial condition for HDE model is not available. In this case one 
%can assume the perturbations of matter component as a source to generate 
%HDE perturbations and use adiabatic condition to find the amplitude of 
%such perturbations. On the other hand, 
%since the energy density of HDE 
%is defined according to event horizon IR cut-off, one can consider the 
%density perturbations of HDE come from the perturbations of the future 
%event horizon \cite{Li:2008zq}. In this case, it has been shown that the adiabatic sound speed of 
%HDE model is positive and therefore the perturbations are stable (see \cite{Li:2008zq}). The nature of 
%HDE perturbations in this case is completely different from the previous fluid description of HDE. In this paper we consider  the HDE as a cosmic fluid which its initial perturbations come from the adiabatic condition.}
%Consequently, if we allow DE clustering then one would expect that
%the DE perturbations affect the growth rate of
%matter perturbation as well
%\citep{Abramo2007,Abramo2009a,Batista2013,Mehrabi:2014ema}.
 As a matter of fact, the key parameter in order to describe the clustering
of DE is the so called
effective sound speed $c_{\rm eff}^2=\delta p_{\rm d}/\delta\rho_{\rm d}$.
Usually, in the literature one can find the following two cases:
(i) homogeneous DE with
$c_{\rm eff}^2=1$ (in units of the speed of light)
and (ii) inhomogeneous DE with $c_{\rm eff}^2=0$.
In the former case the sound horizon is equal or
larger than the Hubble horizon which means that DE
perturbations are taking place only at very large scales.
On the other hand, in case (ii), the sound horizon is much smaller
than the Hubble radius and thus DE perturbations
can grow within the framework of
gravitational instability in a similar manner to matter perturbations
\cite{ArmendarizPicon:1999rj,ArmendarizPicon:2000ah,Garriga:1999vw,Akhoury:2011hr}.
The scenario of DE clustering has been widely investigated in the
literature
\cite{Erickson:2001bq,Bean:2003fb,Hu:2004yd,Ballesteros:2008qk,
dePutter:2010vy,Sapone:2012nh,Batista:2013oca,
Dossett:2013npa,Basse:2013zua,Batista:2014uoa,Pace:2014taa,Steigerwald:2014ava}.
Although it is difficult
to directly measure the amount of DE clustering, there are some indications
that the homogeneous DE has some problems towards reproducing
the observed concentration parameter of the massive galaxy
clusters \cite{Basilakos:2009mz}.
In the context of the spherical collapse model, it has
been shown that inhomogeneous DE models fit better the growth
data than the homogeneous DE scenarios
\cite{Mehrabi:2014ema,Basilakos:2014yda,Nesseris:2014mfa}

Following the above lines, in this work we provide a comprehensive
investigation of the HDE cosmological model at the
background and perturbation levels respectively.
The paper is organized as follows: in section
(\ref{sect:growth}) we first present the main ingredients of the
HDE cosmologies. Then we study the
growth matter perturbations in clustered HDE models and we discuss
the variants from the homogeneous case.
In section (\ref{sect:constraints}), we implement a joint likelihood
analysis involving the latest cosmological data including those of
growth in order to put constraints on the corresponding
cosmological parameters. The growth index of the HDE models is determined
for the first time in section IV.
Finally we summarize our results in section
(\ref{conclude}).

%Indeed, the analysis of cosmological data of type Ia
%supernova \citep{Alam2004,Gong2005a,Gong2005b,Huterer2005,Wang2005}
%show that a time varying DE model results in a better fit comparing
%with the standard cosmological constant.

\section{Growth of perturbations in HDE cosmologies}\label{sect:growth}
Initially, we provide a
brief discussion of the HDE cosmological model in the framework of
Freidmann-Robertson-Walker (FRW) metric and then we derive the basic
differential equations which guide the evolution of
matter (with $P_{m}=0$) and DE
perturbations. Phenomenologically, HDE 
can be treated as an effective dark energy fluid which means that 
one can use the continuity equation 
[see Eq.(\ref{eqn:contdt}) with $P_{\rm d}={\rm w}_{\rm d} \rho_{\rm d}$]. 
Such a description is widely used in this kind of studies
(see for example \cite{Myung:2007pn} and references therein).

\subsection{HDE model}
In the context of the flat FRW metric, the dynamics of the Universe
containing pressure-less matter,
radiation and DE fluids is given by

\begin{equation}\label{eqn:fh3}
 H^{2}=\frac{8\pi G}{3}(\rho_{\rm m}+\rho_{\rm r}+\rho_{\rm d})\;,
\end{equation}
where $H$ is the Hubble parameter, $\rho_{\rm m}$, $\rho_{\rm r}$
and $\rho_{\rm d}$ are the matter, radiation \footnote{When we deal
with the growth of matter DE fluctuations we neglect the radiation
component because we are well inside the matter era.} and DE energy
densities, respectively. Considering that interactions do not take
place among the cosmic fluid components one may write the following
continuity equations
%In HDE
%cosmologies, the energy density of DE is given by
%Eq.~(\ref{eqn:h1}). The energy densities of dust matter, radiation
%and DE are read separately by the following continuity equations
\begin{eqnarray}
&&\dot{\rho}_{\rm m}+3H\rho_{\rm m}=0, \label{eqn:contmt}\\
&&\dot{\rho}_{\rm r}+4H\rho_{\rm r}=0, \label{eqn:contrt}\\
&&{}\dot{\rho}_{\rm d}+3H(1+\rm{w_ d})\rho_{\rm
d}=0\;,\label{eqn:contdt}
\end{eqnarray}
where the dot is the derivative with respect to cosmic time and
$\rm{w_ d}$ is the DE equation of state (hereafter EoS) parameter.
%In the
%equations (\ref{eqn:contmt}, \ref{eqn:contrt} and \ref{eqn:contdt}),
%we assume there is no direct interactions between the cosmic fluids.
Differentiating
Eq.(\ref{eqn:fh3}) and using at the same time
Eq.~(\ref{eqn:h1}),
the continuity equations
(\ref{eqn:contmt}, \ref{eqn:contrt}, \ref{eqn:contdt}) and the
expression $\dot{R}_{\rm h}=1+HR_{\rm h}$,
%and also the expression for
%the energy density $\rho_{\rm d}=3c^2M_{Pl}^2R_{\rm h}^{-2}$
we can obtain the EoS parameter
$\rm{w_d}$ of the HDE model
\begin{equation}\label{eqn:eos}
{\rm w_ d}(z)=-\frac{1}{3}-\frac{2\sqrt{\Omega_{\rm d}(z)}}{3s}\;,
\end{equation}
where $\Omega_{\rm d}(z)=1-\Omega_{\rm m}(z)$ is the
dimensionless energy density of the DE fluid and $z$ is the redshift.
At late enough times, due to the fact that
DE dominates the cosmic expansion of the universe,
we have $\Omega_{\rm d} \to 1$
which means that ${\rm w_ d} \to -\frac{1}{3}-\frac{2}{3s}$.
Within this framework, if we suppose that ${\rm w_{d}}(z_{\star})=-1$
at the special epoch of $z=z_{\star}$ then the value $s$ satisfies
$s=\sqrt{\Omega_{\rm d}(z_{\star})}$. At the present epoch,
$z_{\star}=0$, and for $\Omega_{\rm d0}=\Omega_{\rm d}(0)=0.70$
we compute $s\simeq 0.83$.
%Therefroe, in order to get a phantom like EoS parameter
%$w_{\rm d}<-1$ we need to impose $c<1$, while the quintessence regime
%can be recovered for $c \geq 1$.
On the other hand, at large redshifts $z\gg 1$ ($\Omega_{\rm d} \to 0$)
it is easy to check that
the asymptotic value of the EoS parameter is ${\rm w}_{\infty} \to -1/3$.

%At the late time DE dominated Universe ($\Omega_{\rm
%d}\rightarrow 1$), one can get the phantom like equation of state
%$w_{\rm d}<-1$ for HDE model with $c<1$. On the other hand, the
%quintessence regime $-1\leq \rm{w_ d}<-1/3$ can be achieved in the
%case of $c\geq1$. Some recent observations favor DE models with
%$w_{\rm d}$ crossing the phantom line $-1$ in the near past
%\citep{Zhao2012,Alam2004,Gong2005a,Gong2005b,Huterer2005,Wang2005}.

\begin{figure}
\centering
\includegraphics[width=0.5\textwidth]{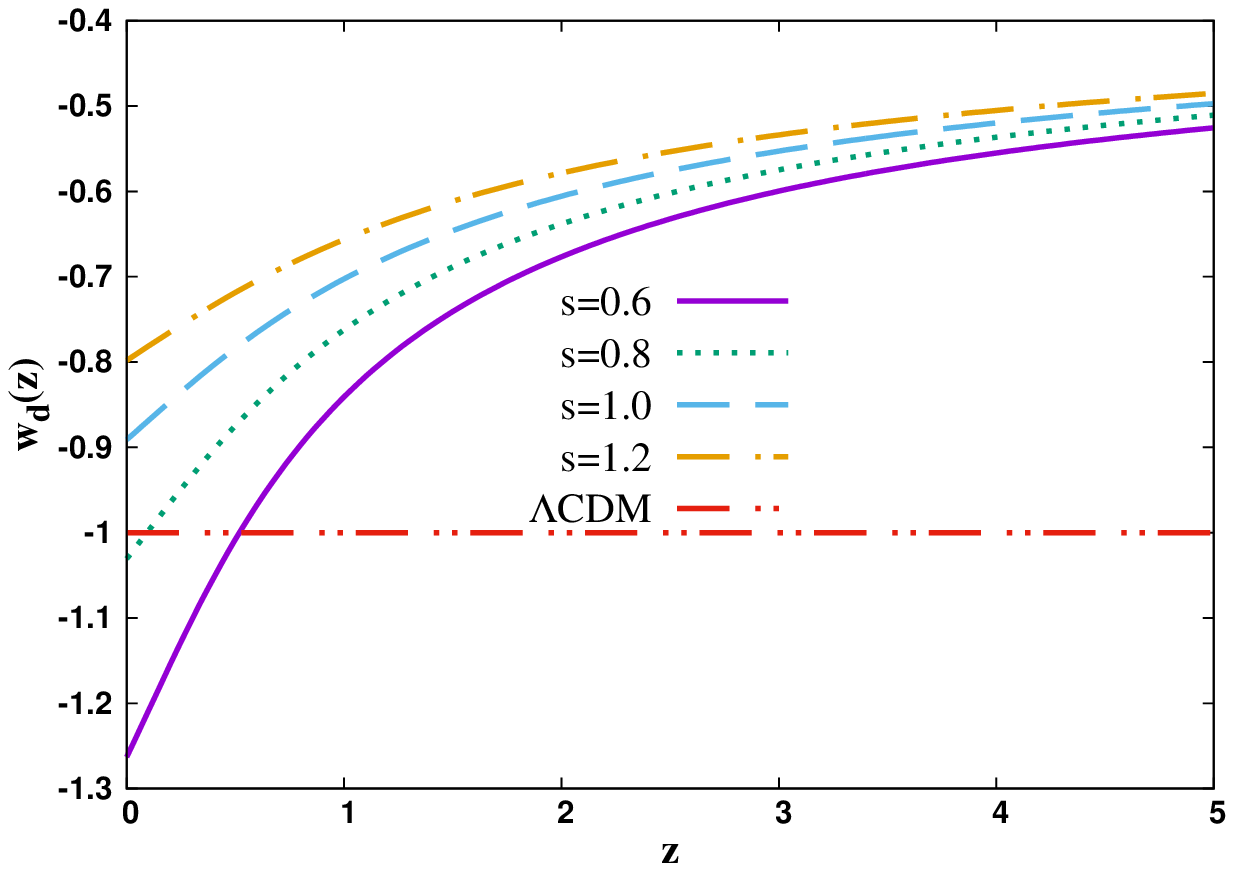}
\includegraphics[width=0.5\textwidth]{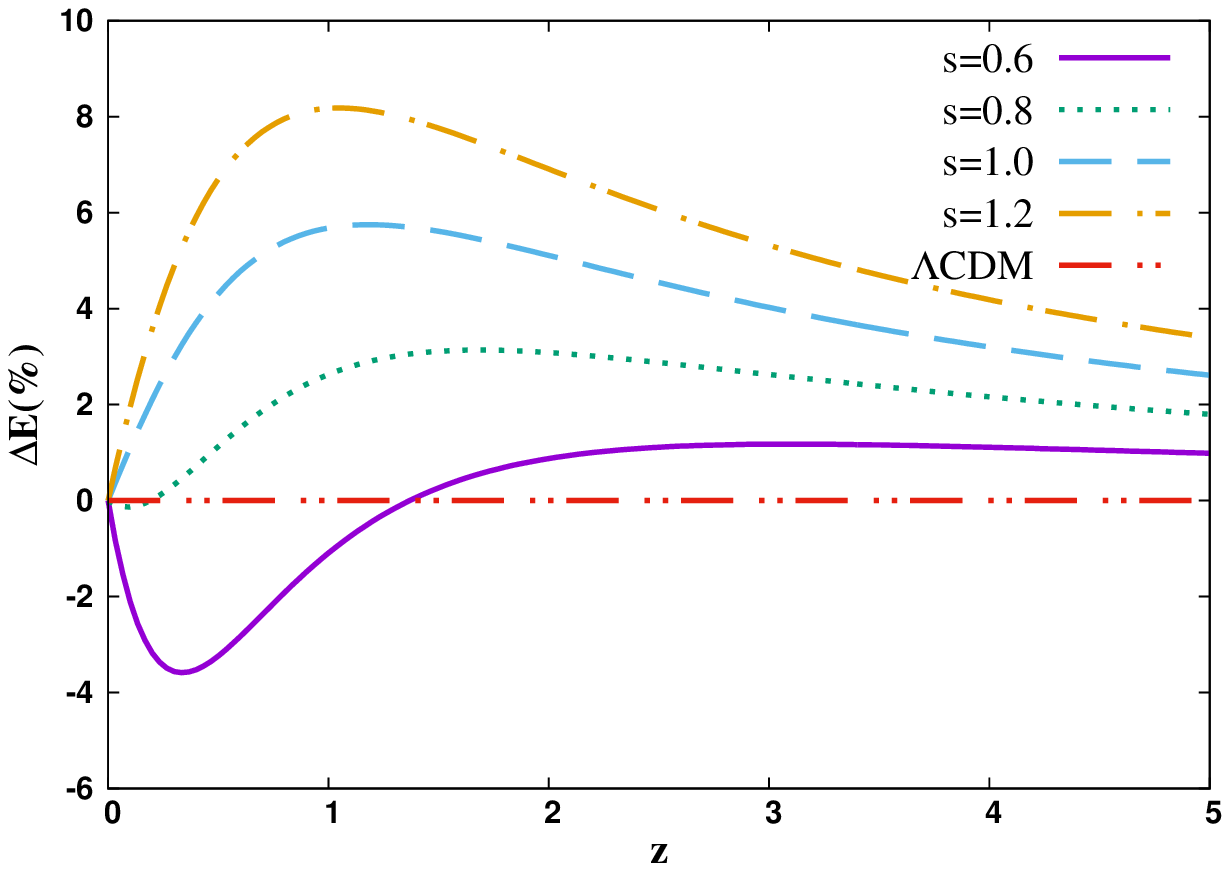}
\includegraphics[width=0.5\textwidth]{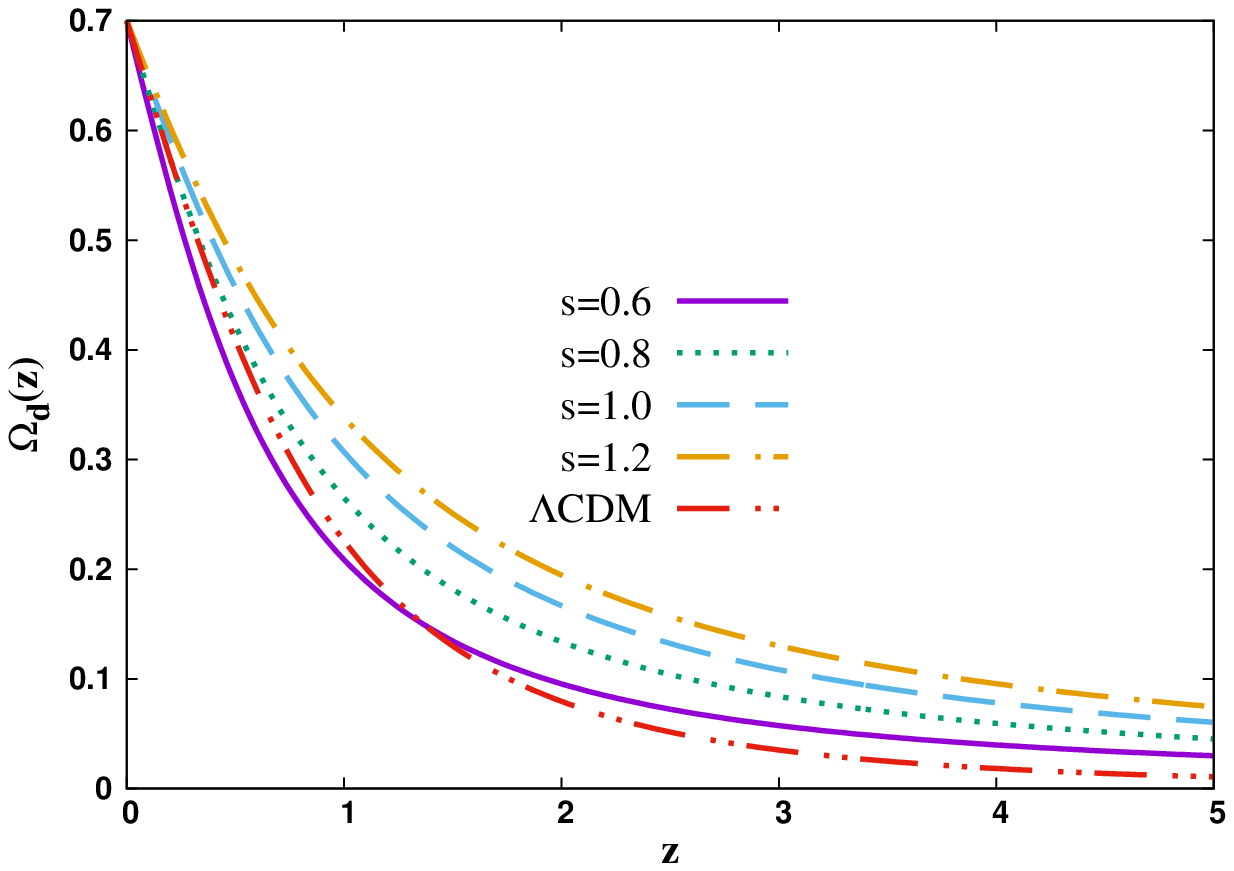}
\caption{{\it Top panel:} Evolution of the equation of state parameter of
HDE model $\rm{w_ d}$ as a function of cosmic redshift $z$ for
different values of the parameter $s$. {\it Middle panel:}
The relative deviation $\Delta E(z)$ of the normalized Hubble parameter for the
HDE models with respect to $\Lambda$CDM.
{\it Bottom panel:} The evolution of
the DE density parameter $\Omega_{\rm d}$.
The dotted-dashed, dashed, dotted and solid curves
correspond to HDE models with $s=1.2$, $s=1$, $s=0.8$ and $s=0.6$ respectively.
Notice, that the reference $\Lambda$CDM model is shown by
 dashed-double-dotted line. In all cases we use
$\Omega_{\rm d0}=1-\Omega_{\rm m0}=0.7$.} \label{fig:back}
\end{figure}

% In HDE models the density parameter $\Omega_{\rm
% d}$ varies with cosmic time $t$.
Now, taking the time derivative of $\Omega_{\rm
d}=\rho_{\rm d}/\rho_{\rm s}=1/(HR_{\rm h})^2$, we can obtain the
evolution of $\Omega_{\rm d}$ in HDE models as follows
\begin{equation}\label{eqn:evol}
a\Omega_{\rm d}^{\prime}=-\frac{\rm{w_ d}}{3}\Omega_{\rm
d}(1-\Omega_{\rm d})\;,
\end{equation}
where the prime is the derivative with respect to scale factor $a$. In
terms of cosmic redshift $z=1/a-1$, Eq.(\ref{eqn:evol}) can
be written as
\begin{equation}\label{eqn:evolz}
\frac{d\Omega_{\rm d}}{dz}=\frac{\rm{w_ d}(z)\Omega_{\rm
d}(z)}{3(1+z)}\Big[1-\Omega_{\rm d}(z)\Big] \;.
\end{equation}

Using the Freidmann equation (\ref{eqn:fh3}) and the continuity
equations (\ref{eqn:contmt}, \ref{eqn:contrt}), we can easily derive
the evolution of the dimensionless Hubble parameter, $E(z)=H(z)/H_0$
(where $H_0$ is the Hubble constant), as follows:
\begin{equation}\label{eqn:hub1}
E^2(z)=\frac{\Omega_{\rm m0}(1+z)^3+\Omega_{\rm
ro}(1+z)^4}{1-\Omega_{\rm d}(z)}\;.
\end{equation}
Obviously, equations (\ref{eqn:eos}),
(\ref{eqn:evolz}) and (\ref{eqn:hub1}) form a system whose
solution provides the evolution of the main cosmological
functions, namely $E(z)$, ${\rm w}_{\rm d}(z)$ and $\Omega_{\rm d}(z)$.
%We solve the coupled system of equations (\ref{eqn:eos}),
%(\ref{eqn:evolz}) and (\ref{eqn:hub1}) in order to obtain the
%dynamics of Hubble flow in HDE models.
As an example, in the upper panel of Fig. (\ref{fig:back}), we
present the evolution of the EoS parameter ${\rm w_ d}(z)$ for the
following HDE models $s=0.6$ (dashed line), $s=0.8$ (dotted line)
$s=1$ (solid line) and $s=1.2$ (dot-dashed line). Notice, that all
models are fixed to $\Omega_{\rm d 0} = 0.7$ at the present time.
%Since the latter functions are directly affected by the value of $c$,
%as an example we select the following three values
%$c=0.6$, $c=1.0$ and $c=1.2$.
%one can assume that the dynamics of the
%expanding Universe is determined by model parameter $c$ of the HDE
%models.
%Since both density parameter of DE,
%  $\Omega_{\rm d}(z)$, and the
%equation of state, $\rm{w_d}$,
%are directly related to the parameter $c$,
%one can assume that the dynamics of the
%expanding Universe is determined by model parameter $c$ of the HDE
%models.
%Here we choose three different values for model parameter
%$c$ as $c=0.6$, $c=1.0$ and $c=1.2$.
%In figure (\ref{fig:back}), we present the
%evolution of the EoS parameter $\rm{w_ d}$ (top panel).
As expected, for $s=0.8$ we reach the phantom regime
prior to the present time.
In the case of $s=0.6$ the EoS parameter crosses the phantom
line $\rm{w_d}=-1$
at the epoch of $z\sim 0.5$, while
for $s=1$ and $s=1.2$ it remains in quintessence regime for each $z$.

Since the Hubble expansion affects the growth of matter
perturbations it is important to understand the behavior of the
Hubble parameter in HDE cosmologies. In this context, we can
appreciate in the middle panel of Fig. (\ref{fig:back}) the relative
difference $\Delta E(z)$ of the normalized Hubble parameters $E_{\rm
HDE}(a)$ with respect to the $\Lambda$CDM solution $E_{\rm \Lambda
CDM}(z)$
\begin{equation}\label{eq:delta-Hubble}
\Delta E(z)=100\times\Big[\frac{E(z)_{\rm HDE}}{E(z)_{\rm \Lambda
CDM}}-1\Big]\;.
\end{equation}
For the quintessence
HDE models ($s=1$ and $s=1.2$), we observe that the quantity
$\Delta E(z)$ is positive for all redshifts which means that the corresponding
cosmic expansion is larger than that of the concordance $\Lambda$CDM model.
There is a visible deviation from the latter around the epoch
$z\sim 0.7$. This deviation becomes at the level of $\sim +5\%$
and $\sim +8\%$ for $s=1$ and $s=1.2$ respectively.
On the other hand, in the case of phantom HDE models (${\rm w_{d}}<-1$)
we have a mixed situation.
Specifically, using $s=0.6$ we can see that at $z\sim 0.7$
the relative difference becomes at the level of $\sim -3.5\%$, while
at high enough redshifts $z\gg 1$ we find that
$E_{\rm HDE}(z)$ deviates form $E_{\rm \Lambda CDM}(z)$ only by $\sim +1\%$.
For $s=0.8$ the maximum relative difference is $\sim +2.5\%$ at
$z\sim 1$, while prior to the present epoch since ${\rm w_{d0}} \sim -1$
we have $E_{\rm HDE}(z=0) \to E_{\rm \Lambda CDM}(z=0)$.

Lastly, in the bottom panel of Fig. (\ref{fig:back}) we plot
$\Omega_{\rm d}$ as a function of redshift. For the majority 
of the HDE models we find $\Omega_{\rm
d}(z)>\Omega_{\Lambda}(z)$. In the case of $s=0.6$ the amount of
DE is a bit higher (lower) than that of $\Lambda$CDM model at high
(low) redshifts.

\subsection{growth of perturbations in HDE models}
Here we briefly discuss the main properties of the linear
perturbation theory within the framework of HDE cosmologies.
Following the general approach
of \citep{Abramo:2008ip} %in pseudo-Newtonain cosmology
the basic equations that describe the evolution of matter and DE
perturbations are given by
\begin{eqnarray}
&&\dot{\delta}_{\rm m} + \frac{\theta_{\rm m}}{a} = 0,\; \label{eq:line1} \\
&&\dot{\delta}_{\rm d} + (1+\rm{w_ d})\frac{\theta_{\rm d}}{a}+3H(c_{\rm eff}^2-w_{\rm d})\delta_{\rm d}= 0,\; \label{eq:line2}\\
&&\dot{\theta}_{\rm m} +H\theta_{\rm m} -\frac{k^2\phi}{a}= 0,\; \label{eq:line3} \\
&&\dot{\theta}_{\rm d} +H\theta_{\rm d}-\frac{k^2c_{\rm eff}^2\delta_{\rm
d}}{(1+\rm{w_ d})a} -\frac{k^2\phi}{a}= 0.\; \label{eq:line4}
\end{eqnarray}
where $c_{\rm eff}^{2}$ is the effective sound speed.
It is well known that at sub-horizon scales
one can extract the scale $k$ from the above equations by
utilizing the Poisson equation (see \citep{Mehrabi:2015hva} and references therein)
\begin{equation}\label{eq:poisson}
-\frac{k^2}{a^2}\phi=\frac{3}{2}H^2[\Omega_{\rm m}\delta_{\rm m}+(1+3c_{\rm eff}^2)\Omega_{\rm d}\delta_{\rm d}]\;,
\end{equation}
The amount of DE clustering depends on the magnitude of its
effective sound speed $c_{\rm eff}^2$ and for
$c_{\rm eff}^2=0$ DE clusters in a similar manner to dark matter. However,
due to the presence of the DE pressure one may expect
that the amplitude of the DE perturbations
is relatively low with respect to that of dark matter.
Notice, that bellow we set $c_{\rm eff}^2=0$.
In the current work we treat DE as a
perfect fluid \footnote{For an imperfect DE fluid see \cite{Bean:2003fb}.}
which implies that
the effective sound speed coincides with the adiabatic sound speed
\begin{equation}\label{eq:c_a}
 c_{\rm a}^2=\rm{w_ d}-\frac{a \rm{w^{\prime}_d}}{3(1+\rm{w_d})}.\;
\end{equation}
where prime denotes derivative with respect to the scale factor,
${\rm w}^{\prime}_{\rm d}=d{\rm w_{d}}/da$.
%but for a imperfect fluid DE, effective sound speed depends on the scale $k=\frac{2\pi}{\lambda}$ and given by \citep{Bean:2003fb}
%\begin{equation}\label{eq:c_eff}
% \frac{\delta p_{\rm d}}{\delta\rho_{\rm d}}=c_{\rm eff}^2^2+3aH(1+\rm{w_ d})(c_{\rm eff}^2^2-c_{\rm a}^2)
% \frac{\theta}{\delta}\frac{1}{k^2}\;,
%\end{equation}
%where $\delta p_{\rm d}$ and $\delta\rho_{\rm d}$  denote first order perturbation of pressure and density of DE, respectively,
% and $\theta$ is the divergence of proper velocity.

Now eliminating $\theta$ from the system of equations
(\ref{eq:line1}, \ref{eq:line2}, \ref{eq:line3} and \ref{eq:line4})
and using $\frac{d}{dt}=aH\frac{d}{da}$ we obtain after some calculations
the following second order differential equations which
describe the evolution of matter and DE perturbations respectively:
\begin{eqnarray}
&\delta''_{\rm m}& + A_{\rm m}\delta'_{\rm m}+
B_{\rm m}\delta_{\rm m}  =  \frac{3}{2a^2}(\Omega_{\rm m}\delta_{\rm m}+\Omega_{\rm d}\delta_{\rm d})\;, \label{eq:sec-ord-delta_m}\\
&\delta''_{\rm d}& + A_{\rm d}\delta'_{\rm d}+
B_{\rm d}\delta_{\rm d}  =  \frac{3}{2a^2}(1+\rm{w_ d})(\Omega_{\rm m}\delta_{\rm m}+\Omega_{\rm d}\delta_{\rm d})\;,  \label{eq:sec-ord-delta_d}
\end{eqnarray}
where the coefficients are
\begin{eqnarray}\label{eq:cof}
A_{\rm m} & = & \frac{3}{2a}(1-\Omega_{\rm d}\rm{w_ d})\;,\\ \nonumber
B_{\rm m} & = & 0\;, \\ \nonumber
A_{\rm d} & = & \frac{1}{a}\left[-3{\rm w_ d}-\frac{a\rm{w'_ d}}{1+\rm{w_ d}}+
\frac{3}{2}(1-\Omega_{\rm d}\rm{w_ d})\right]\;,\\ \nonumber
B_{\rm d} & = & \frac{1}{a^2}\left[-a\rm{w'_ d}+\frac{a\rm{w'_ d}\rm{w_ d}}{1+\rm{w_ d}}-\frac{1}{2}\rm{w_ d}(1-3\Omega_{\rm d}\rm{w_ d})\right]\;,\\ \nonumber
\end{eqnarray}

%It should be noted that to drive these equations we consider the
%full clustering case of DE $(c_{\rm eff}^2^2=0)$.

We would like to stress that Eqs.(\ref{eq:sec-ord-delta_m})
and (~\ref{eq:sec-ord-delta_d}) are both valid in post-Newtonian
and General Relativity (GR) formalisms respectively \citep{Mehrabi:2015hva}.
In order to measure the evolution of DE and matter fluctuations we numerically
integrate the aforementioned equations
from $a=0.001$ till the present time $a=1$ ($z=0$).
Regarding the initial conditions the situations is as follows.
We use the conditions of \citep{Batista:2013oca}:
\begin{eqnarray}
&\delta'_{\rm mi}&=\frac{\delta_{\rm mi}}{a_i},\; \label{eq:delmp_ini}\\
&\delta_{\rm di}&=\frac{1+\rm{w_ d}}{1-3\rm{w_d}}\delta_{\rm mi}, \label{eq:deld_ini}\;\\
&\delta'_{\rm di}&=\frac{4\rm{w'_ d}}{(1-3\rm{w_ d})^2}\delta_{\rm mi}+\frac{1+\rm{w_ d}}{1-3\rm{w_ d}}\delta'_{\rm mi}, \label{eq:deldp_ini}\;
\end{eqnarray}
where we set $\delta_{\rm mi}=1.5\times10^{-4}$ which guarantees
that matter perturbations are in the linear regime ($\delta_{\rm m0}
\ll 1$). Here we focus on the following scenarios: (i) the
holographic DE remains homogeneous ($\delta_{\rm d}=0$) and only the
corresponding matter component clusters (non-clustering holographic
dark energy hereafter NCHDE) and (ii) clustered HDE assuming that
the whole system fully clusters (matter and HDE), namely $c_{\rm
eff}^2=0$ (full clustering  holographic dark energy: FCHDE).

\begin{figure}
\centering
\includegraphics[width=0.5\textwidth]{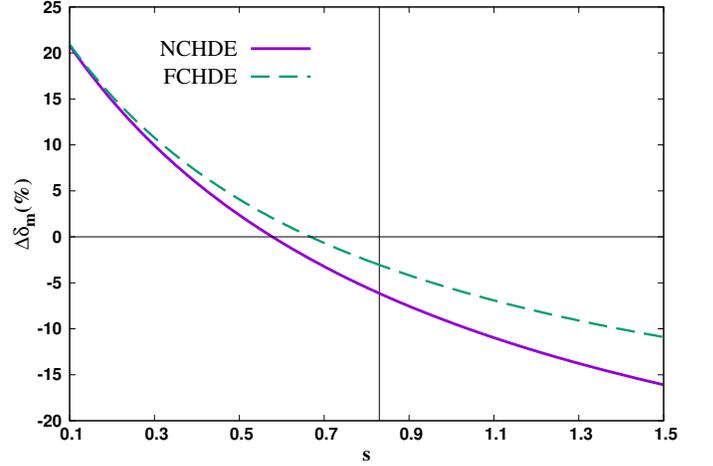}
\caption{The relative difference
$\Delta \delta_{\rm m}(\%)$ at the present epoch versus $s$.
The solid (dashed) lines stands for NCHDE (FCHDE) model.
The vertical line indicates the $w_{\rm d}=-1$ ($s=0.83$
for $\Omega_{\rm d0}=0.7$) line.}\label{fig:delm}
\end{figure}

In figure (\ref{fig:delm}) we show the relative deviation
of the HDE matter fluctuations (at the current epoch)
with respect to those of $\Lambda$CDM as a function of $s$
\begin{equation}\label{eq:deltam}
\Delta \delta_{\rm m} = 100 \times \left[\frac{(\delta_{\rm m0})_{\rm HDE}}{(\delta_{\rm m0})_{\Lambda{\rm CDM}}} - 1 \right].\;
\end{equation}
Notice, that the vertical line at $s=0.83$ separates the phantom
from the quintessence region [see discussion after Eq.(\ref{eqn:eos})].
%. Indeed for $c\le 0.83$ one may check that
%the EoS parameter at the present obeys $w_{\rm d 0}\le -1$, while
%the opposite holds for $c>0.83$.

First we observe that $\Delta \delta_{\rm m}$ is a decreasing function of $s$
and it lies in the interval $(-20\%,20\%)$.
These differences imply
that the deviations of the HDE matter fluctuations depend on the
initial assumptions and limitations imposed in the general system of
equations (\ref{eq:line1}), (\ref{eq:line2})
and (\ref{eq:delmp_ini})-(\ref{eq:deldp_ini}).
%that govern the matter and DE perturbations.
Second we verify that for small values of $s$ the
Hubble friction ($H\delta_{\rm m}$) is small
with respect to that of $\Lambda$CDM (see also figure \ref{fig:back}),
which means that the corresponding matter fluctuations in NCHDE/FCHDE
models are larger than those of the concordance $\Lambda$ cosmology.
The opposite is true for large values of $s$.
Also, we find that the relative deviations
between NCHDE and FCHDE models are negligible when $s<0.3$.
For $s>0.3$ the term $\Omega_{\rm d}\delta_{\rm d}$ in Eq. (\ref{eq:line1})
starts to get power
and thus we have $\delta_{\rm m}^{(\rm FCHDE)} > \delta_{m}^{(\rm NCHDE)}$.

%growth of dust perturbations in NCHDE
% (FCHDE) model is larger than the concordance $\Lambda$CDM Universe.
% Opposingly, for $c>0.6$ ($c>0.68$) the growth of structures in
% NCHDE (FCHDE) model slower than $\Lambda$CDM Universe. This result
% is considerable because for small values of $c$ the Hubble friction
% is smaller and consequently the growth of perturbations becomes
% greater. On the other hand, in the cases of high values of $c$ the
% Hubble parameter becomes larger and the growth of perturbations
% slows down

\begin{figure}
\centering
\includegraphics[width=0.5\textwidth]{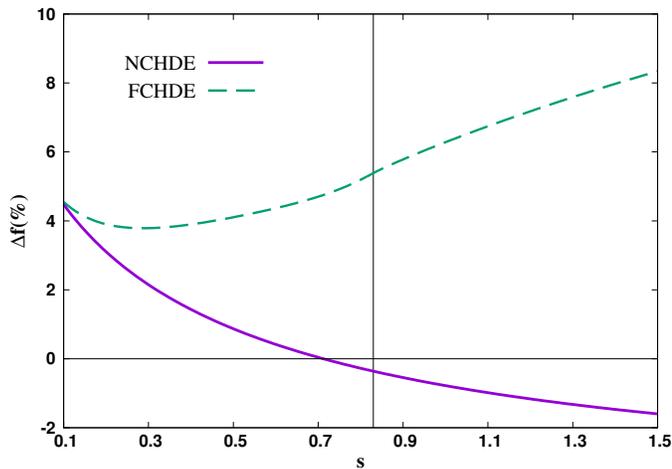}
\caption{The relative difference of the growth rate
of clustering at present time as a function of $s$.
%relative to $\Lambda$CDM model.
The style of lines can be found in figure (\ref{fig:delm}).
The vertical line represents the phantom border,
namely $s=0.83$ for $\Omega_{\rm d0}=0.7$.}\label{fig:delf}
\end{figure}

Moreover, we calculate the growth rate of clustering
in HDE models and compare it with concordance $\Lambda$CDM  model.
The growth rate function is given by $f(a)=d\ln{\delta_{\rm m}}/d\ln{a}$
[see Eq.(\ref{fzz221}) below].
In this case the corresponding relative difference is given by
\begin{equation}\label{eq:deltaf}
\Delta f= 100 \times \left(\frac{f_{\rm HDE}}{f_{\Lambda{\rm CDM}}}
- 1 \right).\;
\end{equation}
In figure (\ref{fig:delf}) we plot the quantity
$\Delta f(\%)$ at the present epoch as a function of $s$. Obviously,
for the FCHDE model the growth rate of matter perturbations is always
larger than $\Lambda$CDM. As an example, prior to the
phantom line ($s\sim 0.83$) we find a $\sim 4\%$ difference.
In the case of NCHDE model we see that $\Delta f(\%)$
becomes positive (or negative) for $s<0.75$ (or $s>0.75$) and the
relative difference is $\sim 0.4\%$ at $s\sim 0.83$.

\begin{figure}
\centering
\includegraphics[width=0.5\textwidth]{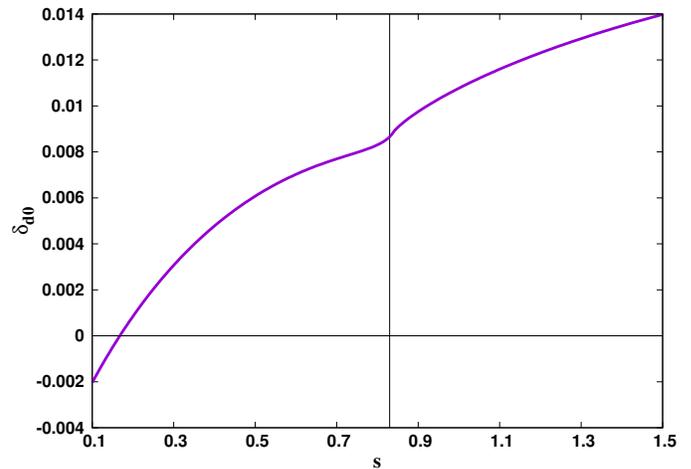}
\caption{DE perturbations at present time as a function of $s$.
The vertical line presents the phantom border.}\label{fig:deld}
\end{figure}

Finally, in figure (\ref{fig:deld}) we plot $\delta_{\rm d0}$ versus $s$ and
we observe that the current DE perturbations increase, due to the term  $1+{\rm w_d}$
in equation (\ref{eq:sec-ord-delta_d}), as a
function of $s$. To this end,
close to $s\simeq 0.83$ we find that $\delta_{\rm d0} \simeq 0.008$.
%For $c<0.2$
%, DE perturbations is negative, which is due to phantom like
%equation of state of HDE model. Moreover, present time value of DE
%perturbations is larger by increasing the value of $c$. Such
%behavior expected from equation (\ref{eq:sec-ord-delta_d}), due to
%the factor $1+{\rm w_d}$ in the source term. For example for
%quintessence case (${\rm w_d}<-1$), smaller value of ${\rm w_d}$
%(large values of $c$) results a higher value of source term and
%leads to larger DE perturbations.

\section{Holographic models versus data}\label{sect:constraints}
In this section we attempt to compare
the HDE models against the latest observational data.
Specifically, we perform an overall statistical analysis using the
geometrical data (SnIa \citep{Union2.1:2012}, BAO \citep{Beutler:2011hx,Padmanabhan:2012hf,Anderson:2012sa,Blake:2011en},CMB \citep{Hinshaw:2012aka}, Big Bang Nucleosynthesis
\cite{Serra:2009yp,Burles:2000zk}, $H(z)$ data \cite{Moresco:2012jh,Gaztanaga:2008xz,Blake:2012pj,Anderson:2013zyy}) and the growth data as gathered by
\cite{Mehrabi:2015hva}.

%\textbf{I added the CMB data from WMAP data release 9 and commented a paragraph here}
%Concerning the CMB shift parameter the situation is as
%follows. This standard ruler is
%somewhat model dependent especially to those DE models which
%either contain massive neutrinos or when the EoS parameter varies strongly with
%redshift \cite{Elgaroy:2006tp,Corasaniti:2007rf}.
%The performace of the shift parameter has been studied in
%\cite{Elgaroy07}.
%As far as the HDE scenarios are concerned, it
%is easy to check [see Eq. (\ref{eqn:eos}) and Top panel of Fig.1] that
%the corresponding EoS parameter changes from ${\rm w_d}=-\frac{1}{3}$
%in the past to ${\rm w_d}=-\frac{1}{3}-\frac{2}{3c}$
%(for $c=1$ ${\rm w_d}=-1.3$) in the far future,
%and thus we decide to exclude the CMB shift parameter from the
%statistical analysis in order to avoid possible biases.

We would like to point that the geometrical data, the growth data, the
covariances, the joint $\chi_{\rm tot}^{2}({\bf p})$ function including
that of the Akaike information criterion\footnote{ $N/n_{\rm fit}>0$,
the Akaike information criterion is given by AIC$=\chi^{2}_{\rm min}+2n_{\rm fit}$,
where $n_{\rm fit}$ is the number of free parameter \cite{Akaike:1974}.}
and the MCMC algorithm are presented
in \cite{Mehrabi:2015hva}. The above statistical vector contains
the corresponding cosmological parameters, namely
${\bf p}=\{\Omega_{\rm DM},\Omega_{\rm b},h,s,\sigma_{8}\}$,
where $\Omega_{\rm DM}$ and $\Omega_{\rm b}$ are the dimensionless
energy densities of
pressure-less dark matter and baryons, respectively and $\sigma_8$ is
the variance of matter perturbations within the sphere of $R=8{\rm
h^{-1}Mpc}$ at present time. We would like to mention that the total
$\Omega_{\rm m}(a)$ is written as
$\Omega_{\rm m}(a)\equiv \Omega_{\rm DM}(a)+\Omega_{\rm b}(a)$.

\begin{figure*}
\centering
\includegraphics[width=\textwidth]{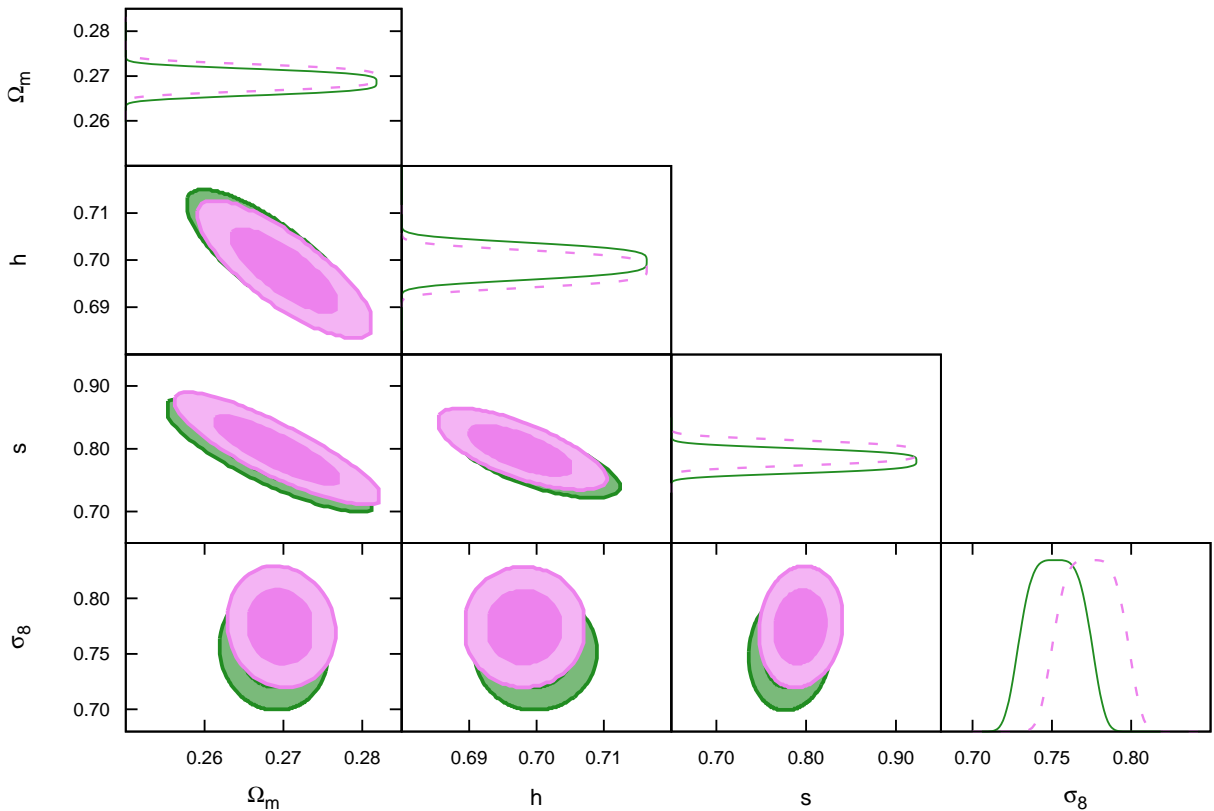}
\caption{1$\sigma$ and 2$\sigma$ confidence regions for FCHDE and NCHDE models. Green background
 (violet foreground) area is for FCHDE (NCHDE) model.  The likelihood function for FCHDE (NCHDE) has been shown by solid (dashed) curve. } \label{fig:cont}
\end{figure*}

\begin{table}
 \begin{center}
 \setlength{\extrarowheight}{10pt}
  \begin{tabular}{| c | c |c |}
   \hline
   Parameters          & HDE & $\Lambda$CDM \\
   \hline
   $\Omega_{\rm m0}$ & $0.267^{+0.0019+0.0026}_{-0.0016-0.0028}$ &$0.276^{+0.0017+0.0029}_{-0.0015-0.0027}$\\
   \hline
   $h$                 & $0.696^{+0.005+0.007}_{-0.004-0.007}$ &$0.70^{+0.004+0.008}_{-0.005-0.009}$\\
   \hline
    $s$               & $0.782^{+0.035+0.079}_{-0.033-0.075}$ &$-$\\
   \hline
      $\rm{w_{d0}}$       & $-1.062$ &$-1.$\\
   \hline
  \end{tabular}
    \caption{The best fit values of parameters using the geometry measurements data set for HDE and concordance $\Lambda$CDM models.  $\rm{w_{d0}}$ indicate the EoS at present time.}
 \label{tab:res1}
 \end{center}
\end{table}

\begin{table}
\setlength{\extrarowheight}{10pt}
  \begin{tabular}{| c | c |c |}
   \hline
   Parameters          & NCHDE & FCHDE \\
      \hline
   $\Omega_{\rm m0}$ & $0.269^{+0.0025+0.0034}_{-0.0023-0.0038}$ &$0.268^{+0.0024+0.0031}_{-0.0021-0.0036}$\\
   \hline
   $h$                 & $0.693^{+0.004+0.006}_{-0.003-0.005}$ &$0.695^{+0.003+0.006}_{-0.003-0.005}$\\
   \hline
   $s$               & $0.792^{+0.027+0.064}_{-0.025-0.068}$ &$0.780^{+0.026+0.069}_{-0.028-0.066}$\\
   \hline
    $\sigma_{\rm 8}$         & $0.776^{+0.041+0.079}_{-0.043-0.078}$ & $0.752^{+0.040+0.075}_{-0.043-0.077}$\\
   \hline
      $\rm{w_{d0}}$      & $-1.051$ &$-1.063$\\
      \hline
  \end{tabular}
    \caption{The best fit values of parameters using the joint analysis of geometry measurements
    data + growth rate data for NCHDE and FCHDE models.}
 \label{tab:res2}
\end{table}

The observational constraints are summarized in Table I and II,
where for the former case we utilize only the geometrical data while
for the latter case we have included the growth data in the
overall likelihood analysis. Specifically, we find the following.

In the case of the geometrical data,

\begin{itemize}
\item for the HDE model: $\chi^2_{\rm min}=588.1$, $n_{\rm fit}=4$  and
AIC=$596.1$;
\item for the $\Lambda$CDM model, $\chi^2_{\rm min}=588.9$, $n_{\rm fit}=3$  and
AIC$_{\Lambda}$=$594.9$.
\end{itemize}

In the case of geometrical and growth rate data,

\begin{itemize}
\item for the NCHDE model, $\chi^2_{\rm min}=595.8$, $n_{\rm fit}=5$ and
AIC=$605.8$;

\item For FCHDE model, $\chi^2_{\rm min}=595.2$, $n_{\rm fit}=5$ and AIC=$605.2$;

\item for the $\Lambda$CDM model, $\chi^2_{\rm min}=595.9$, $n_{\rm fit}=4$
and AIC$_{\Lambda}$=$603.9$
\end{itemize}

The above results show that in all possible cases
$\Delta {\rm AIC}=|{\rm AIC}-{\rm AIC}_{\Lambda}| \le 2$ which implies
that the observational data are consistent with the current
cosmological models (for a relevant
discussion see Ref.\cite{Mehrabi:2015hva}).
Also in figure (\ref{fig:cont}) we present the 1$\sigma$ and 2$\sigma$
combined likelihood contours for the explored holographic DE models
(FCHDE: green region and NCHDE: violet region). We observe that
both HDE models (FCHDE and NCHDE) provide the same statistical results
within 1$\sigma$ uncertainties.

\begin{figure}
\centering
\includegraphics[width=0.5\textwidth]{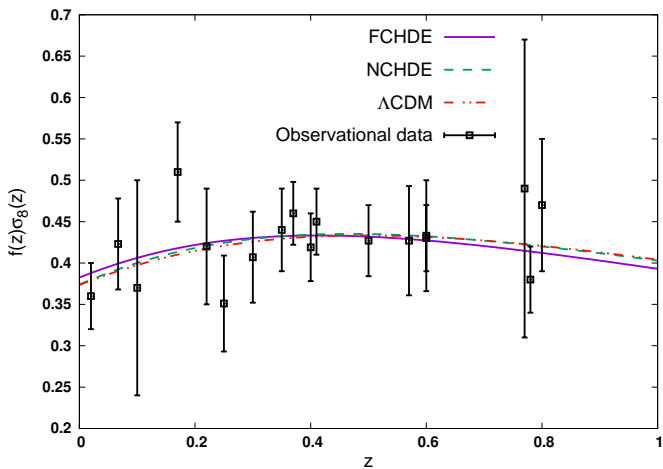}
\caption{The growth rate function of matter perturbations using the best fit values of cosmological parameters. The violet solid, green
dashed and red dotted-double-dashed curves show the  FCHDE, NCHDE and $\Lambda$CDM
 models, respectively. Observational data have been shown by open square with their error bars (see Table \ref{tab:fsigma8data}). }\label{fig:fs8}
\end{figure}

To this end, using the aforesaid cosmological parameters
in figure (\ref{fig:fs8}) we compare
the observed (open boxes) with the theoretical evolution of the
growth rate $f\sigma_{8}(z)$. Notice, that
$f(z)$ is the rate of the growing mode (see next section),
$\sigma_{8}(z)=\sigma_{8}D(z)$ and $D(z)$ is the growth factor
normalized to unity at the present time.
As is expected from the AIC analysis
the current DE models provide almost the same growth rate predictions.

\begin{center}
\begin{table}
\caption{The $f\sigma_8(z)$ data points including their references and surveys.}
\begin{tabular}{ccc}
\hline
\hline
z & $f\sigma_8(z)$ & Reference \\
\hline
$0.02$  & $0.360\pm0.040$ & \cite{Hudson:2012gt}\\
$0.067$ & $0.423\pm0.055$ & \cite{Beutler:2012px}\\
$0.10$  & $0.37\pm0.13$   & \cite{Feix:2015dla}\\
$0.17$  & $0.510\pm0.060$ & \cite{Percival:2004fs}\\
$0.35$  & $0.440\pm0.050$ & \cite{Song:2008qt,Tegmark:2006az}\\
$0.77$  & $0.490\pm0.180$ & \cite{Guzzo:2008ac,Song:2008qt}\\
$0.25$  & $0.351\pm0.058$ & \cite{Samushia:2011cs}\\
$0.37$  & $0.460\pm0.038$ & \cite{Samushia:2011cs}\\
$0.22$  & $0.420\pm0.070$ & \cite{Blake:2011rj}\\
$0.41$  & $0.450\pm0.040$ & \cite{Blake:2011rj}\\
$0.60$  & $0.430\pm0.040$ & \cite{Blake:2011rj}\\
$0.60$  & $0.433\pm0.067$ & \cite{Tojeiro:2012rp}\\
$0.78$  & $0.380\pm0.040$ & \cite{Blake:2011rj}\\
$0.57$  & $0.427\pm0.066$ & \cite{Reid:2012sw}\\
$0.30$  & $0.407\pm0.055$ & \cite{Tojeiro:2012rp}\\
$0.40$  & $0.419\pm0.041$ & \cite{Tojeiro:2012rp}\\
$0.50$  & $0.427\pm0.043$ & \cite{Tojeiro:2012rp}\\
$0.80$  & $0.47\pm0.08$   & \cite{delaTorre:2013rpa}\\
\hline
\hline
\end{tabular}
\label{tab:fsigma8data}
\end{table}
\end{center}

\section{The Growth index in Holographic cosmology}
Let us focus now on the analysis of the
growth index of matter perturbations $\gamma$. It is well known
that the growth rate of clustering is given in terms
$\gamma$ and $\Omega_{m}(a)$
(see Refs.~\cite{Peebles1993,Silveira:1994yq,Wang:1998gt,Linder:2003dr,Linder:2004ng,Linder:2007hg,Nesseris:2007pa,Lue:2004rj})
 \be
\label{fzz221}
f(a)=\frac{d{\rm ln} \delta_{\rm m}}{d{\rm ln} a}\simeq \Omega^{\gamma}_{m}(a) \;.
\ee
Basically, for a given cosmological model
the growth index of matter fluctuations provides a characteristic
identity at the perturbation level.
As an example, within the context of GR it has been shown that
the asymptotic value of the growth index
is $\gamma_{\infty}=\frac{3({\rm w}-1)}{6{\rm w}-5}$, where in this case the dark energy
EoS parameter is constant
\cite{Silveira:1994yq,Wang:1998gt,Linder:2004ng,Linder:2007hg,Nesseris:2007pa}.
Of course, for the usual $\Lambda$CDM model (${\rm w=-1}$) the above
formula boils down to $\gamma_{\infty}^{(\Lambda)} =6/11$.
Notice, that for some specific types of modified gravity
models we refer the reader to
\cite{Linder:2004ng,Linder:2007hg,Wei:2008ig,Gong:2008fh,Fu:2009nr,Gannouji:2008wt,Tsujikawa:2009ku,Basilakos:2013ij}.

Now following the methodology of
Abramo et al. \cite{Abramo:2007iu,Abramo:2008ip}
we write the following equation
\be
\label{ooodedelta}
\ddot{\delta}_{\rm m}+ 2H\dot{\delta}_{\rm m}=
\frac{3H^{2}}{2}\left[\Omega_{\rm m}\delta_{\rm m}+\Omega_{\rm d}\delta_{d}(1+3c_{\rm eff}^2)\right].
\ee
Changing the variables from $t$ to $a$
($\frac{d\delta_{\rm m}}{dt}=aH\frac{d\delta_{\rm m}}{da}$)
we obtain
\begin{equation}\label{odedelta}
a^{2}\delta_{\rm m}^{\prime \prime}+
a\left(3+\frac{\dot{H}}{H^2}\right)\delta_{\rm m}^{\prime}=
\frac{3}{2}\left[\Omega_{\rm m}\delta_{\rm m}+(1+3c_{\rm eff}^2)
\Omega_{\rm d}\delta_{\rm d}\right]\;,
\end{equation}
where
\begin{equation}\label{eos22}
\frac{{\dot H}}{H^{2}}=\frac{d{\rm ln} H}{d{\rm ln}a}
=-\frac{3}{2}-\frac{3}{2}{\rm w}_{\rm d}(a)\Omega_{\rm d}(a)\;,
\end{equation}
and $\Omega_{\rm d}(a)=1-\Omega_{\rm m}(a)$.
Of course, for $c_{\rm eff}^2=0$ (FCHDE model) Eq.(\ref{odedelta}) reduces to
Eq.(\ref{eq:sec-ord-delta_m}) as it should. We would like to point
that by definition in the case of the $\Lambda$CDM model
DE perturbations vanish, $\delta_{\rm d}\equiv 0$.

Furthermore, based on the above and utilizing the first equality
of Eq.(\ref{fzz221})
it is easy to prove that
\begin{equation}\label{fzz222}
\frac{df}{d{\rm ln}a}+
\left(2+\frac{d{\ln}H}{d{\rm ln}a}\right)f+f^{2}
=\frac{3\mu \Omega_{m}}{2}\;,
%\frac{df}{d{\rm ln}a}+f^{2}+\left(\frac{1}{2}-\frac{3}{2}w_{\rm d}\Omega_{\rm d}\right)f
% =\frac{3\mu \Omega_{\rm m}}{2}
\end{equation}
or
\begin{equation}\label{Poll}
-(1+z)\frac{d\gamma}{dz}{\rm ln}(\Omega_{\rm m})+\Omega_{\rm m}^{\gamma}+
3{\rm w}_{\rm d}\Omega_{\rm d}\left(\gamma-\frac{1}{2}\right)+\frac{1}{2}=
\frac{3}{2}\Omega_{\rm m}^{1-\gamma}\mu\;.
\end{equation}
where we have inserted $f(z)=\Omega_{m}(z)^{\gamma(z)}$, Eq.(\ref{eos22}) and
$\frac{df}{da}=-(1+z)^{-2}\frac{df}{dz}$ in Eq.(\ref{fzz222}).
Also, the quantity $\mu(a)$, which characterizes the status of the HDE,
is given by 
\begin{equation} \label{VV}
\mu(a)=\left\{ \begin{array}{cc} 1
\;\;
       &\mbox{Homogeneous HDE}\\
  1+\frac{\Omega_{\rm d}(a)}{\Omega_{\rm m}(a)}\Delta_{\rm d}(a)(1+3c_{\rm eff}^2)
\;\;
       & \mbox{Clustered HDE}
       \end{array}
        \right.
\end{equation}
where
%\begin{equation}
%\mu(a)=1+\frac{\Omega_{\rm d}(a)}{\Omega_{\rm m}(a)}\Delta_{\rm d}(a)(1+3c_{\rm eff}^2)
%\end{equation}
%and 
$\Delta_{\rm d}\equiv \delta_{\rm d}/\delta_{\rm m}$.
Notice, that below we use the abbreviation CHDE which corresponds to
clustered holographic dark energy model with $c_{\rm eff}^2\ne 0$. Of course,
for $c_{\rm eff}^2=0$ we recover the fully clustered case 
(FCHDE model see section II).

It has been proposed that an approximated solution of Eq.(\ref{Poll})
%the above non-linear differential equation
is written as a first order Taylor expansion around the present epoch
$a(z)=1$ (see Refs. \cite{Ishak:2009qs},\cite{Belloso:2011ms},\cite{DiPorto:2011jr},\cite{Pouri:2014nta,Basilakos:2012uu,Basilakos:2012ws,Wu:2009zy})
\begin{equation}
\label{gzzz}
\gamma(a)=\gamma_{0}+\gamma_{1}\left[1-a(z)\right]\;.
\end{equation}
where $a(z)=1/(1+z)$. 
Furthermore, evaluating Eq.(\ref{Poll}) for
$z=0$ and with the aid of Eq.(\ref{gzzz}) we can write the coefficient
$\gamma_{1}$ as a function of $(\Omega_{\rm m0},\gamma_{0},{\rm w}_{\rm d0},\mu_{0})$ (see
also Ref.\cite{Polarski:2007rr})
\begin{equation}
\label{Poll2}
\gamma_{1}=\frac{\Omega_{\rm m0}^{\gamma_{0}}+3{\rm w}_{d0}(\gamma_{0}-\frac{1}{2})
\Omega_{\rm d0}+\frac{1}{2}-\frac{3}{2}\Omega_{\rm m0}^{1-\gamma_{0}} \mu_{0}}
{{\rm ln}  \Omega_{\rm m0} }\;,
\end{equation}
where $\mu_{0}=\mu(z=0)$ and ${\rm w}_{d0}={\rm w}_{\rm d}(z=0)$.

In order to obtain the evolution of the growth index (\ref{gzzz}) 
we need to know the value of $\gamma_{0}$. It is easy to check 
from Eq.(\ref{gzzz})
that $\gamma_{\infty}\simeq \gamma_{0}+\gamma_{1}$ at large redshift $z\gg 1$ and 
thus $\gamma_{0} \simeq \gamma_{\infty}-\gamma_{1}$.
 %Since $\gamma_{0}$ is unknown, in order to proceed with the
%analysis we need to introduce an additional constrain. 
%One may check
%from Eq.(\ref{gzzz})
%that $\gamma_{\infty}\simeq \gamma_{0}+\gamma_{1}$ at large redshift $z\gg 1$.
Therefore, it is important 
to calculate the asymptotic value of the
growth index from first principles. 

Fortunately, Steigerwald et al. \cite{Steigerwald:2014ava} 
developed a general mathematical approach
which provides $\gamma_{\infty}$ analytically
[see Eq.(8) in \cite{Steigerwald:2014ava} and the discussion 
in \cite{Basilakos:2015vra}] 
for a large family of DE models. In particular,
based on Steigerwald et al. \cite{Steigerwald:2014ava} one can use
\be
\label{g000}
\gamma_{\infty}=\frac{3(M_{0}+M_{1})-2(H_{1}+N_{1})}{2+2X_{1}+3M_{0}}
\ee
where the following quantities have been defined:
\be \label{Coef1}
M_{0}=\left. \mu \right|_{\omega=0}\,,
\ \
M_{1}=\left.\frac{d \mu}{d\omega}\right|_{\omega=0}
\ee
and
\be \label{Coef2}
N_{1}=0\,,\ \
H_{1}=-\frac{X_{1}}{2}=\frac{3}{2}\left. {\rm w_{\rm d}}(a)\right|_{\omega=0}
\,.
\ee
%$M_{0}=[\mu]_{\omega=0}$, $M_{1}=[\frac{d\mu}{d\omega}]_{\omega=0}$,
%$N_{1}=0$ and $H_{1}=-\frac{X_{1}}{2}=\frac{3}{2}[{\rm w_{d}}(a)]_{\omega=0}$.
%(for more details see appendix A).
Notice, that in the notation of Steigerwald et al. \cite{Steigerwald:2014ava} 
the basic cosmological functions are provided in terms of the variable
%that we set 
$\omega={\rm ln}\Omega_{\rm m}(a)$. The latter implies that
for $z\gg 1$ we have $\Omega_{\rm m}(a)\to 1$
[or $\Omega_{\rm d}(a)\to 0$] and thus $\omega \to 0$.

Bellow we provide the main results of the above analysis.

\subsection{Homogeneous HDE: NCHDE model}
Considering the non-clustering holographic
dark energy (NCHDE) model, $\mu(a)=1$, we find 
[see Eqs.(\ref{Coef1}), (\ref{Coef2})]
$$
\{ M_{0},M_{1},H_{1},X_{1}\}=\{ 1,0,\frac{3{\rm w}_{\infty}}{2},-3{\rm w}_{\infty}\}
$$
where we have used $w_{\infty}\equiv w_{\rm d}(a)_{\omega=0}$.
If we substitute the above coefficients into Eq.(\ref{g000}) then
the asymptotic growth index is given by
\be
\gamma_{\infty} =\frac{3({\rm w}_{\infty}-1)}{6{\rm w}_{\infty}-5} .
\ee
As expected, we recover the standard
$\Lambda$CDM value $\gamma_{\infty}^{(\Lambda)}=6/11$
for ${\rm w}_{\infty}=-1$. In the case of holographic cosmology, namely
$w_{\infty}=-1/3$ we find $\gamma_{\infty}^{(\rm NCHDE)}=4/7$
which is somewhat
larger ($\sim 4.8\%$) than that of the concordance $\Lambda$ cosmology.

\begin{figure}[t]
\includegraphics[width=0.5\textwidth]{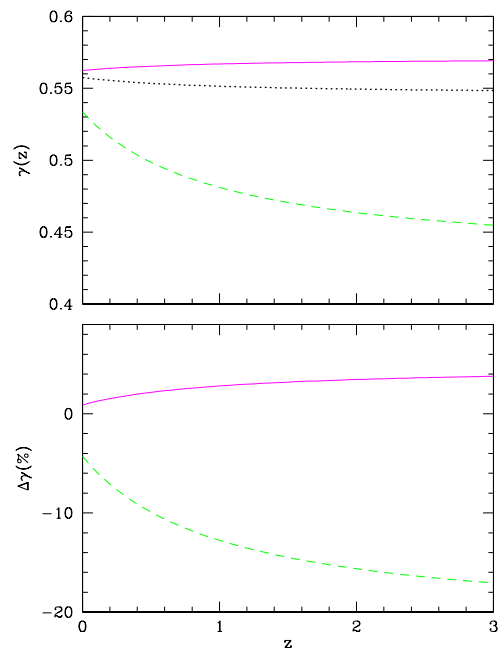}
\caption{{\em Upper Panel:} The growth index of matter
perturbations (\ref{gzzz}) as a function of $z$.
The lines are as follows. The results for
NCHDE (solid curve) and FCHDE models are given by
the solid and dashed curves. Also, the thin dotted line
corresponds to the concordance $\Lambda$ cosmology.
{\em Bottom Panel:} The relative difference
$[1-\gamma(z)/\gamma^{(\Lambda)}(z)]\%$ of the growth index
for the NCHDE and FCHDE models with respect to
$\Lambda$CDM.}
\end{figure}

Knowing the value of $\gamma_{\infty}$,
inserting the expression $\gamma_{0} \simeq \gamma_{\infty}-\gamma_{1}$ into
Eq.(\ref{Poll2}) and using the best fit values of the cosmological parameters
(see Table II) we find $(\gamma_{0},\gamma_{1})^{(\rm NCHDE)}\simeq (0.562,0.09)$.
In the case of the $\Lambda$CDM we obtain
$(\gamma_{0},\gamma_{1})\simeq (0.557,-0.012)$.
In the upper panel of figure 5 we plot the growth index evolution (\ref{gzzz})
for the NCHDE (solid curve) and $\Lambda$CDM models (thin dotted curve)
respectively.
We observe that the growth index evolution of the NCHDE model is somewhat
larger than the $\Lambda$CDM
cosmological model. Specifically, we find that the corresponding
relative deviations (see bottom panel of Fig.5) are
$[1- \gamma^{(\rm NCHDE)}/ \gamma^{(\Lambda)}]
\sim 3\%$.

\subsection{Clustered HDE: CHDE model}
Now we concentrate on the clustered holographic DE in which
the quantity $\mu(a)$ is given by the second branch of Eq.(\ref{VV}).
First of all we need to define the functional form of $\Delta_{\rm d}$.
Based on Eqs.(\ref{eqn:eos}) and (\ref{eq:deld_ini})
we can write
\be
\Delta_{\rm d}=\frac{1+{\rm w_{d}}}{1-3{\rm w_{d}}}=
\frac{s-\sqrt{\Omega_{\rm d}}}{3s+3\sqrt{\Omega_{\rm d}}}=
\frac{s-\sqrt{1-{\rm e}^{\omega} }}{3s+3\sqrt{1-{\rm e}^{\omega} }}
%\frac{c-\sqrt{\Omega_{\rm d}}{3c+3\sqrt{\Omega_{\rm d}}}=
%\frac{c-\sqrt{1-{\rm e}^{\omega}}{3c+3\sqrt{1-{\rm e}^{\omega}}}
\ee
where ${\rm e}^{\omega}=\Omega_{\rm m}(a)=1-\Omega_{\rm d}(a)$.
Under the latter conditions $\mu(\omega)$ is written as
\be
\mu(\omega)=1+(1+3c_{\rm eff}^2)\left(\frac{1-{\rm e}^{\omega}}{\rm e^{\omega}}\right)
\left(\frac{s-\sqrt{1-{\rm e}^{\omega} }}{3s+3\sqrt{1-{\rm e}^{\omega} }}\right) .
\ee
Therefore, from Eqs.(\ref{Coef1}) and (\ref{Coef2}) it is easy to show that
$$
\{ M_{0},M_{1},H_{1},X_{1}\}=\{ 1,-\frac{(1+3c_{\rm eff}^2)}{3},\frac{3{\rm w}_{\infty}}{2},-3{\rm w}_{\infty}\}
$$
and for ${\rm w}_{\infty}=-1/3$ we get via Eq.(\ref{g000})
\be \label{g001}
\gamma^{(\rm CHDE)}_{\infty}=\frac{3(1-c_{\rm eff}^2)}{7} \;.
\ee

Obviously, if we set $c_{\rm eff}^2=0$
in the above equation then
we provide the asymptotic value of the 
fully clustered HDE (FCHDE) model,
$\gamma^{(\rm FCHDE)}_{\infty}=3/7$. 
Finally, concerning the evolution of the growth index the situation
is as follows. Substituting
$\gamma_{0} \simeq \gamma^{(\rm FCHDE)}_{\infty}-\gamma_{1}$ into
Eq.(\ref{Poll2}) and utilizing the cosmological parameters
of Table II
we get $(\gamma_{0},\gamma_{1})^{(\rm FCHDE)}\simeq (0.534,-0.105)$.
The dashed curve 
in the upper panel of figure 5 corresponds to 
the evolution of $\gamma(z)$ for the FCHDE model (dashed curve). 
In this case we verify that 
the growth index 
strongly deviates with respect to that of the usual $\Lambda$ cosmology.
For example close to the present epoch
the departure can be of the order of
$[1- \gamma^{(\rm FCHDE)}/ \gamma^{(\Lambda)}]
\sim -4\%$ while at relative large redshifts $z\sim 1$ we get
$[1- \gamma^{(\rm FCHDE)}/ \gamma^{(\Lambda)}]
\sim -13\%$.

%Finally, we would like to finish this section by presenting 
%two special CHDE models.
%Specifically, if we select $c_{\rm eff}^2=-3/11$ then this particular CHDE model
%provides $\gamma_{\infty}^{(\rm CHDE)}=6/11$ [see Eq.(\ref{g001})] which is 
%equal to that of $\Lambda$CDM model.
%On the other hand, if we impose $c_{\rm eff}^2=-1/3$ into Eq.(\ref{g001}) 
%then we have $\gamma^{(\rm CHDE)}_{\infty}=4/7$. 
%Practically
%this special CHDE model is cosmologically
%equivalent with that of the
%NCHDE (homogeneous) 
%model because they share exactly the same cosmic expansion and matter
%perturbations, despite the fact that the former cosmological
%scenario allows DE clustering.

\section{Conclusion}\label{conclude}

%While majority of the dynamical DE models are presented in the
%phenomenological frameworks, in this work we focused on the
%theoretically justified model the so-called holographic dark energy
%(HDE) model originated from the quantum gravity theory
%\cite{Susskind:1994vu}. 

In this work we investigated the performance of the holographic dark energy
(HDE) model which is originated from the 
holographic principle in quantum gravity theory
\cite{Susskind:1994vu}. 
%In particular, according to the holographic
%Dynamical DE models not only alleviate theoretical 
%problem of $\Lambda$CDM scenario but 
%also permit DE fluctuations in similar fashion to dark matter
%\cite{Abramo:2007iu,Abramo:2008ip,Batista:2013oca,Mehrabi:2014ema}.
First we studied the growth of matter perturbations in clustered HDE
models and discussed the differences from the homogeneous case. 
Second we performed a joint likelihood analysis using the 
latest observational data in order to place tight 
constraints on the cosmological parameters. 
Lastly, we provided (for the first time) the growth index of matter 
perturbations for the homogeneous and inhomogeneous HDE models. 
Specifically, the main results of the current 
work are summarized as follows:

(i) We find that for $s<0.6$ 
the amplitude of matter perturbations $\delta_{\rm m}$ 
of both
clustered and homogeneous HDE models are larger than
those of the concordance $\Lambda$ cosmology, while 
the opposite holds for $s>0.6$. Notice, that $s$ is a constant 
which is related with the holographic DE density [see Eq. \ref{eqn:h1} ]. 
Moreover, the relative deviation between clustered and
homogeneous HDE cosmologies remains small in the case 
of phantom HDE, while there are differences in 
the case of quintessence HDE models (see Fig. \ref{fig:delm}).
As far as the growth rate 
of clustering, $f(a)$, is concerned we find that for the clustered HDE models
$f(a)$ is always larger than that of the 
concordance $\Lambda$CDM model. In the case of homogeneous HDE models
the growth rate of clustering depends on the value of
$s$, namely $f(a)>f_{\Lambda}(a)$ for $s<0.7$ (the opposite 
holds for $s>0.7$: see Fig. \ref{fig:delf}).

%We observe that in clustered HDE model the perturbations of DE
%enhance the growth function of structures compare to homogenous DE
%models.

(ii) The overall likelihood analysis showed that both clustered and
homogeneous HDE models provide the same cosmological parameters within
1$\sigma$ uncertainties. In this context
based on the AIC analysis we found that 
the HDE models are consistent with the observational data.
%the results of AIC analysis
%indicate that in all cases we have $\Delta {\rm AIC}=|{\rm AIC}-{\rm
%AIC}_{\Lambda}| \le 2$ which means that 
%representing that the observational data are
%consistent with both clustered and non-clustered HDE models. In particular, we found the best value of parameters for both models provide a phantom like EoS at present time. In addition the best value of $\sigma_8$ at present time for NCHDE is slightly larger than FCHDE model.  

(iii) Finally, we focused our analysis on the growth index of 
matter fluctuations. Assuming that the holographic 
dark energy is homogeneous then 
the asymptotic value of the growth index of matter perturbations
is given by $\gamma \approx 4/7$ which 
differs by $\sim 4.8\%$ from that of 
$\Lambda$CDM, $\gamma^{(\Lambda)} \approx 6/11$. 
The situation is different in the case of inhomogeneous holographic 
dark energy. Within this framework, 
we investigated the growth index and we verified that it is 
strongly affected by the 
dark energy perturbations. In particular, we found 
$\gamma \approx 3(1-c_{\rm eff}^2)/7$, where $c_{\rm eff}^2$ is the effective sound speed.
Finally, assuming that the dark energy is allowed to cluster
in a similar fashion to dark matter ($c_{\rm eff}^2=0$)
we find that the asymptotic value of 
the growth index ($\gamma \approx 3/7$)
is strongly affected by the DE perturbations.
Such an effect can be used, especially in the light of
the next generation of surveys \cite{Sapone:2013wda},
in order to test whether the DE
perturbations appear in the observed universe.
In other words, if one is able to show that the observed growth index is
close to $\sim 3/7$ then this is a hint that holographic
DE perturbations exist in nature.

%=================================================================

 \bibliographystyle{apsrev4-1}
  \bibliography{ref}

\end{document}